\newcommand{\bB}{\mathbf{B}}
\newcommand{\bv}{\mathbf{v}}

\newcommand{\nn}{\nonumber}

\newcommand{\nb}{\nabla}

       \newcommand{\Dc}{ {\mathcal{D}} }
       
       \newcommand{\Fc}{ {\mathcal{F}} }
       \newcommand{\Gc}{ {\mathcal{G}} }

       \newcommand{\Mc}{ {\mathcal{M}} }
       \newcommand{\Nc}{ {\mathcal{N}} }

       \newcommand{\Uc}{ {\mathcal{U}} }

\documentclass[twocolumn,amsmath,amssymb,aps]{revtex4-1}

\usepackage[utf8]{inputenc}
\usepackage[T1]{fontenc}
\usepackage{amsmath}

\usepackage{mathtools}
\usepackage{amsfonts}
\usepackage{graphicx}
\usepackage{dcolumn}
\usepackage{bm}
\usepackage[export]{adjustbox}
\usepackage{wrapfig}
\usepackage{lipsum}
\usepackage[titletoc,toc,title]{appendix}
\usepackage[colorlinks]{hyperref}
\usepackage{color}

\begin{document}

\title{Ellipticity conditions for the extended MHD Grad-Shafranov-Bernoulli equilibrium equations}

\author{D. A. Kaltsas}
 \email{dkaltsas@cc.uoi.gr}
\author{G. N. Throumoulopoulos}%
 \email{gthroum@uoi.gr}
\affiliation{
Department of Physics, University of Ioannina,\\ GR 451 10 Ioannina, Greece
}%
 \author{P. J. Morrison}
 \email{morrison@physics.utexas.edu}
\affiliation{%
Department of Physics and Institute for Fusion Studies,\\ University of Texas, Austin, Texas 78712, USA 
}

\date{\today}

\begin{abstract}
In this study we find the point of transition between elliptic and hyperbolic regimes for the axisymmetric extended MHD equilibrium equations. The  ellipticity condition is expressed via a single inequality but is more involved than the corresponding two-fluid ones, due to the imposition of the quasineutrality condition,  and is also more complicated than the Hall MHD ellipticity condition, due to electron inertia.  In fact,  the inclusion of electron inertia  is responsible for peculiar results; namely,   even the static equilibrium equations can become hyperbolic. 
\end{abstract}

\pacs{Valid PACS appear here}
\maketitle


The study of the equilibrium properties of fusion and astrophysical plasmas is usually performed within the framework of ordinary, single-fluid magnetohydrodynamics (MHD), which considers the plasma as a single conducting fluid without taking into account the individual contributions of the constituent  species of particles, i.e., the ions, electrons and possibly neutral particles. A better and more accurate description includes the consideration of this coexistence of the various  components of the plasma. 
 The easiest way to observe the effects that emerge due to this coexistence is to perform equilibrium studies within the framework of two-fluid theory. 
There exist various studies in this field and most of them adopt certain assumptions regarding the effect of each fluid or their combined behavior in order to simplify the analysis.  The most generic case is rather difficult and challenging because a complete two-fluid equilibrium study requires the solution of two force balance equations coupled to the Maxwell equations, on account of the long range interactions, and also the consideration of two continuity equations for the particle densities, which in turn are  involved in thermodynamical relations.
 A first assumption that reduces this system is the assumption of quasineutrality, reducing the two continuity equations into one and eliminating the electric field in the resulting force balance equation. This assumption leads to extended MHD (XMHD), which is a quasineutral two-fluid model 
expressed in terms of the total velocity and current density \cite{Lust1959,Kimura2014}. In addition an expansion in the smallness of the electron to ion mass ratio is also performed. Upon neglecting electron inertia the XMHD model reduces to the well known and extensively studied model of Hall magnetohydrodynamics (HMHD). 
 
 In the present  study we show how the quasineutrality condition, although it reduces the number of equations that have to be considered for a fully self-consistent description, inserts a peculiarity into the system of equilibrium equations derived in \cite{Kaltsas2017,Kaltsas2018a}:  the  two flux functions representing  the electron and the ion contributions are connected through a single Bernoulli equation and a single mass density function.  This  feature, that is not a characteristic of the complete two-fluid theory, introduces a complication in deriving ellipticity conditions for the XMHD equilibrium system of equations,  rendering the condition more involved than those for the two-fluid system. However there are special cases where the ellipticity condition is reduced to more convenient forms that indicate interesting conclusions. Such a case is static equilibria, static in the sense that the motion of the electron and the ion fluids are  restricted so as to prevent macroscopic mass flow, in which case we can prove that ellipticity is not always possible, despite the fact that if we neglect electron inertia, the absence of macroscopic flow implies ellipticity, as it is well known in the case of MHD. 


The classification of PDEs and systems of PDEs into elliptic, parabolic and hyperbolic ones, is fundamental in the theory of differential equations (e.g.\ [\onlinecite{Evans2010}]). Boundary value problems (BVPs) with elliptic equations or systems of equations under Dirichlet, Neumann, or Robin boundary conditions are well-posed. On the other hand hyperbolic equations are usually related to evolutionary problems.  Typically ellipticity is defined for systems of linear PDEs (e.g.\ for the specific case of second order systems see \cite{Morrey1953}) because 
it is a property defined pointwise  and is completely depended on the principal symbol of the differential operator;  hence,  the definition can be extended in order to include quasilinear systems as it is done below. Consider a second order  system of $M$ quasilinear partial differential equations in $N$ independent and $M$ dependent variables of the following form:
\begin{eqnarray}
 &&\sum_{j=1}^{M}\sum_{\ell,n=1}^{N}\tau_{ij}^{\ell n}(x,u, u_x)\frac{\partial^2u_j}{\partial x_\ell\partial x_n}\nn\\
 &&\hspace{1cm} -f_i(x,u,u_x)  =0\,, \qquad i=1,...,M \label{pde}
\end{eqnarray}
where $x=(x_1,...,x_N)\in \Dc\subset \mathbb{R}^N$, $u=(u_1,...,u_M)\in \Uc\subset\mathbb{R}^M$, $\tau_{ij}^{\ell n}$ are the coefficients of the second order derivatives in \eqref{pde} and by $u_x$ we denote the first order derivatives of the dependent variables. The classification of the system depends only upon its principal symbol, or characteristic matrix, which for arbitrary real scalars $\lambda=(\lambda_1,...,\lambda_N)$, is defined as
\begin{eqnarray}
\tau[\lambda]=\left[\sum_{\ell,n=1}^{N}\tau_{ij}^{\ell n}(x,u, u_x)\lambda_\ell\lambda_n\right]\,, \label{symbol}
\end{eqnarray}
which is an $M\times M$ matrix with rows and columns labeled by $i$  and $j$,  respectively.

{ \em Definition:} The second order quasilinear system \eqref{pde} is called elliptic if $\forall x\in \Dc$, $det\left(\tau[\lambda]\right)\neq 0$ $\forall \lambda\neq 0$. That is $det\left(\tau[\lambda]\right)$ has to be positive or negative definite $\forall \lambda\neq 0$.

Ellipticity is generally desired for equilibrium studies because they rely on solving boundary value problems, which as stated above, are well-posed and well understood in the elliptic regime. It is also  known that solutions to elliptic equations have no discontinuous derivatives. Such discontinuities are related to  jumps in equilibrium profiles and shock formation, which certainly introduce additional numerical challenges. In ordinary MHD, describing fusion plasmas,   the boundaries   between elliptic and hyperbolic regimes are determined by the magnitude of the poloidal flow. Weak poloidal flows render the equilibrium problem elliptic and thus its solution can be attained by standard methods for boundary value problems; however,  when poloidal flows  have larger magnitudes, then mixed elliptic-hyperbolic regimes, i.e.,  situations for which the equilibrium system is hyperbolic in one part of the domain and elliptic in the other part, emerge.  This implies the existence of discontinuities and jumps in profile quantities such as  the plasma density \cite{Betti2000}.  The connection of strong poloidal sheared flows with the formation of internal transport barriers that are associated with the transition to high confinement modes and whose emergence comes with the formation of steep gradients in equilibrium profiles, establishes a link between mixed elliptic-hyperbolic equilibria with transonic flows and high-mode confinement. 

For the reasons mentioned above it is important to know where the boundaries between elliptic and hyperbolic regimes are located. The ellipticity conditions for single fluid MHD have been derived in several instances e.g. \cite{Hameiri1983,Goedbloed2004,Guazzotto2014}. For the complete two-fluid Grad-Shafranov-Bernoulli equilibrium system, ellipticity conditions are provided in \cite{Goedbloed2004},  while there are analogous conditions for simplified versions, e.g.,  in \cite{Ishida2004} for two-fluid equilibria with massless electrons, in \cite{Hameiri2013a,Hagstrom2014,Ito2007} for the Hall MHD model with scalar and anisotropic electron pressure. 

For reasons of comparison and completeness we give here the well-known ellipticity conditions for axisymmetric MHD and HMHD equations and in addition the respective two-fluid conditions.

 In the context of MHD the axisymmetric Grad-Shafranov-Bernoulli system is elliptic if 
\begin{eqnarray}
0\leq \frac{v_p^2}{v_{Ap}^2}<\frac{c_s^2}{c_s^2+v_A^2}\,,\;\;  v_s^2<v_p^2<v_A^2\,, \nn\\
v_A^2<v_p^2<v_f^2 \,, \label{ellipticity_mhd}
\end{eqnarray} 
where $v_p$ is the poloidal plasma velocity, $c_s$ is the speed of sound, $v_A$ the poloidal Alfv\'en speed, while $v_s$ and $v_f$ correspond to the slow and fast magnetosonic wave speeds, respectively. We can see that within the framework of ordinary MHD there exist two elliptic regions, the second one, which involves stronger flows, is interrupted by the so-called Alfv\'en singularity encountered when the poloidal flow speed coincides with the poloidal Alfv\'en speed. This makes the Grad-Shafranov equation singular and a global equilibrium solution cannot be constructed. It is interesting that the speed of sound is not a transition point,   the transition points being defined by the trailing cusp speed in the wave-front diagram, $c_s^2/(c_s^2+v_A^2)$, and the characteristic speeds of the slow and fast magnetosonic waves.

The ellipticity conditions for two-fluid equilibria acquire a much simpler form and only one elliptic region exists, viz
\begin{eqnarray}
v_{ip}^2<c_{is}^2\,,\quad \mbox{and} \quad v_{ep}^2<c_{es}^2\,, \label{ellipticity_2f}
\end{eqnarray}
where $c_{js}^2=\Gamma p_j/(m_jn_j)$, $j=i,e$ for polytropic gases with adiabatic index $\Gamma$, deduced by reversing the hyperbolicity conditions in \cite{Goedbloed2004}. In the case of Hall MHD the ellipticity condition, derived in \cite{Hameiri2013a}, becomes 
\begin{eqnarray}
v_p^2<c_{s}^2\,, \label{ellipticity_hall}
\end{eqnarray}
where $c_s^2=c_{is}^2+c_{es}^2$ which holds true for HMHD and XMHD due to the quasineutrality condition.
Conditions \eqref{ellipticity_2f} and \eqref{ellipticity_hall} show hydrodynamic behavior within the two-fluid context, with transitions to hyperbolicity when the poloidal speed reaches  the corresponding sound speed.  
One would expect that since the XMHD model is essentially a quasineutral two-fluid model, would exhibit a similar behavior. However as we show below the quasineutrality condition introduces  complication in the XMHD formalism. We reveal this complication by deriving the ellipticity condition for the most generic system of XMHD equilibrium equations and later on we discuss some special cases.

The definition of ellipticity, as given above is clear and allows the classification of systems such as the following, which describes axisymmetric barotropic XMHD equilibria \cite{Kaltsas2018a} in cylindrical coordinates $(r,\phi,z)$: 
\begin{eqnarray}
&&\hspace{-0.cm}(\gamma^2+d_e^2)\Fc'(\varphi)r^2\nb\cdot\left(\frac{\Fc'(\varphi)}{\rho}\frac{\nb\varphi}{r^2}\right)+\frac{\mu}{\gamma-\mu}\Delta^*\psi\nn\\&&\hspace{-0.cm}=\Fc'(\varphi)[\Fc(\varphi)+\Gc(\xi)]+r^2\rho \Mc'(\varphi) -\rho \frac{\varphi-\xi}{(\gamma-\mu)^2}\,, \label{as_gs_1} \\
&&\hspace{-0.cm}(\mu^2+d_e^2)\Gc'(\xi) r^2 \nb\cdot\left(\frac{\Gc'(\xi)}{\rho} \frac{\nb \xi}{r^2} \right)-\frac{\gamma}{\gamma-\mu}\Delta^*\psi\nn \\&&\hspace{-0.cm}=\Gc'(\xi) [\Fc(\varphi)+\Gc(\xi)]+r^2 \rho \Nc'(\xi) +\rho \frac{\varphi-\xi}{(\gamma-\mu)^2}\,, \label{as_gs_2}
\end{eqnarray}
\begin{eqnarray}
&&\hspace{-.5cm}\Delta^*\psi=\frac{\rho}{d_e^2}\left(\psi-\frac{\mu \varphi-\gamma \xi}{\mu-\gamma}\right)\,, \label{as_gs_3} \\
&&\hspace{-.5cm}h(\rho)=\Mc(\varphi)+\Nc(\xi)- \frac{v^2}{2}\nn
\\&&\hspace{2cm}-\frac{d_e^2}{2\rho^2}\left[J_\phi^2+r^{-2}|\nb (rB_\phi)|^2\right]\,,
 \label{bernoulli_gen_ax}
\end{eqnarray}
where $\Delta^*:=r^2\nb\cdot\left(\nb/r^2\right)$ is the elliptic Shafranov operator and $J_\phi=-r^{-1}\Delta^*\psi $ is the toroidal current density   and $\rho$ denotes the mass density.  Note that all quantities are normalized to Alfv\'en units and so $v_A=1$. The functions $\varphi=\psi^*+\gamma r v_\phi$ and $\xi:=\psi^*+\mu r v_\phi$ are related to the poloidal components of the ion and electron fluid flow, respectively; $\gamma:= (d_i + \sqrt{d_i^2 + 4 d_e^2})/2$ and $\mu:= (d_i - \sqrt{d_i^2 + 4 d_e^2})/2$,  where $d_i$ and $d_e$ are the normalized ion and electron skin depths, respectively.  The flux function $\psi^*$ is the poloidal flux function of the generalized magnetic field $\bB^*:=\bB+\nabla\times(\nabla\times\bB/\rho)$. The magnetic field $\bB$ has a toroidal component given by 
\begin{eqnarray}
B_\phi=r^{-1}\left[\Fc(\varphi)+\Gc(\xi)\right]\,. \label{b_phi}
\end{eqnarray}
and a poloidal component $\bB_p=\nabla\psi\times\nabla\phi$. The same decomposition applies for the velocity field with 
\begin{equation}
\bv_p=\rho^{-1}\nb\left(\gamma\Fc+\mu\Gc\right)\times\nb\phi\,.\label{v_p}
\end{equation} 
and 
\begin{equation}
v_\phi=r^{-1}\frac{\varphi-\xi}{\gamma-\mu}\,, \label{v_phi}
\end{equation}
Note that for $d_e=0$ one obtains the axisymmetric Hall MHD equilibrium system \citep{Throumoulopoulos2006}.

For the classification of the system \eqref{as_gs_1}--\eqref{bernoulli_gen_ax} we are interested in knowing the principal symbol, which depends only on the coefficients of second order derivatives of \eqref{as_gs_1}--\eqref{as_gs_3}. An interesting property  of Grad-Shafranov-Bernoulli (GSB) systems such as the system above, is that the second order derivatives in the flux functions are not only those that appear explicitly in the Grad-Shafranov (GS) equations,  but   additional terms coming from the involvement of the mass density $\rho$ in  the differential operators;   according to the Bernoulli equation $\rho=\rho(r,\varphi,\xi,|\nabla\varphi|^2,|\nabla\xi|^2)$, so $\nabla\rho$ will contain second order derivatives. By denoting 
\begin{eqnarray}
\rho':=\frac{\partial\rho}{\partial|\nabla\varphi|^2}\,,\qquad \dot{\rho}:=\frac{\partial\rho}{\partial|\nabla\xi|^2}\,,
\end{eqnarray}
we can rewrite the equilibrium system as follows: 
\begin{eqnarray}
&&(\gamma^2+d_e^2)\frac{\Fc'^2}{\rho r^2}\big[\left(1-\alpha
\varphi_r^2\right)\partial_{rr}\varphi \label{sec_ord_gs_1}\\
&&\hspace{0.5cm}+\left(1-\alpha\varphi_z^2\right)\partial_{zz}\varphi-2\alpha\varphi_r\varphi_z\partial_{rz}\varphi-\beta \varphi_r\xi_r \partial_{rr}\xi\nn \\
&&\hspace{1.5cm}-\beta \varphi_z\xi_z\partial_{zz}\xi-\beta(\varphi_r\xi_z+\varphi_z\xi_r)\partial_{rz}\xi\big]\nn \\
&&\hspace{3cm}+ \mbox{ lower order terms } =0\,,\nn\\
&&(\mu^2+d_e^2)\frac{\Gc'^2}{\rho r^2}\big[\left(1-\beta
\xi_r^2\right)\partial_{rr}\xi \label{sec_ord_gs_2}\\
&&\hspace{0.5cm}+\left(1-\beta\xi_z^2\right)\partial_{zz}\xi-2\beta\xi_r\xi_z\partial_{rz}\xi-\alpha \varphi_r\xi_r \partial_{rr}\varphi\nn \\
&&\hspace{1.5cm}-\alpha \varphi_z\xi_z\partial_{zz}\varphi-\alpha(\varphi_r\xi_z+\varphi_z\xi_r)\partial_{rz}\varphi\big]\nn \\
&&\hspace{3cm}+\mbox{ lower order terms } =0\,,\nn\\
&&\partial_{rr} \psi+\partial_{zz}\psi+\mbox{ lower order terms }  =0\,,\label{sec_ord_gs_3}
\end{eqnarray}
where $\alpha:=2\rho'/\rho$ and $\beta:=2\dot{\rho}/\rho$. Therefore,  according to the definition \eqref{symbol},  the principal symbol of the system \eqref{sec_ord_gs_1}--\eqref{sec_ord_gs_3} is 
\begin{widetext}
\begin{eqnarray}
&&\hspace{-0.5cm}\tau[\lambda_1,\lambda_2]=\nn\\
&&\begin{pmatrix}
C_1\left[(1-\alpha \varphi_r^2)\lambda_1^2+(1-\alpha\varphi_z^2)\lambda_2^2-2\alpha\varphi_r\varphi_z \lambda_1\lambda_2\right] & -C_1\beta\left[\varphi_r\xi_r \lambda_1^2+\varphi_z\xi_z \lambda_2^2+(\varphi_r\xi_z+\varphi_z\xi_r)\lambda_1\lambda_2\right] & 0\\ \\
-C_2\alpha\left[\varphi_r\xi_r \lambda_1^2+\varphi_z\xi_z \lambda_2^2+(\varphi_r\xi_z+\varphi_z\xi_r)\lambda_1\lambda_2\right]& C_2\left[(1-\beta \xi_r^2)\lambda_1^2+(1-\beta\xi_z^2)\lambda_2^2-2\beta\xi_r\xi_z \lambda_1\lambda_2\right] & 0 \\ \\ 
0 & 0 & \lambda_1^2+\lambda_2^2
\end{pmatrix}\!\!,
\end{eqnarray}
where $C_1:=(\gamma^2+d_e^2)\Fc'^2/(\rho r^2)$ and $C_2:=(\mu^2+d_e^2)\Gc'^2/(\rho r^2)$.  The determinant of the characteristic matrix is 
\begin{eqnarray}
det(\tau)(\lambda_1,\lambda_2)=C_1C_2(\lambda_1^2 + \lambda_2^2)^2 \big[\lambda_1^2(1- \alpha \varphi_r^2-\beta \xi_r^2)+\lambda_2^2(1- \alpha \varphi_z^2- \beta \xi_z^2) 
-2\lambda_1\lambda_2(\alpha \varphi_r\varphi_z+\beta \xi_r\xi_z) \big]\nn\\
=:C_1C_2(\lambda_1^2 + \lambda_2^2)^2P(\lambda_1,\lambda_2)\,.
\end{eqnarray}
\end{widetext}
For free functions $\Fc(\varphi)$ and  $\Gc(\xi)$ with  $\Fc'\neq 0$ and $\Gc'\neq 0$ $\forall x\in \Dc$,  the coefficient $C_1C_2$ can be ignored since it is strictly positive.
Clearly for $\Fc',\Gc'\neq 0$  and $\lambda_1,\lambda_2\neq 0$ the determinant can be zero if and only if the homogeneous polynomial $P(\lambda_1,\lambda_2)$ has real roots. Thus the ellipticity condition for XMHD equilibrium equations can be summarized as follows: 
\begin{eqnarray}
P(\lambda_1,\lambda_2)\neq 0\,, \quad \forall \lambda_1,\lambda_2\neq 0\,.
\end{eqnarray}
We can prove, by directly computing the roots of $P(\lambda_1,\lambda_2)$ with respect to either $\lambda_1$ or $\lambda_2$, that no real roots exist if  
\begin{eqnarray}
&&1 -\alpha|\nabla \varphi|^2- \beta |\nabla\xi|^2  \label{ellipt_cond_1} \\
&&\hspace{1 cm} + \alpha \beta \left( |\nabla\varphi|^2|\nabla\xi|^2-(\nabla\varphi\cdot\nabla\xi)^2\right)> 0\,.
\nn
\end{eqnarray}
At this point it remains  to compute $\alpha$ and $\beta$ in terms of the equilibrium quantities. This can be done by performing implicit differentiation of the Bernoulli equation with respect to $|\nabla\varphi|^2$ and $|\nabla\xi|^2$ (e.g.\  see \cite{Hameiri1983,Hameiri2013a}). The final expressions for $\alpha$ and $\beta$ are
\begin{eqnarray}
\alpha&=&\frac{-(\gamma^2+d_e^2)\Fc'^2}{\rho^2 r^2 
\big[c_s^2-v_p^2-\frac{d_e^2}{\rho^2r^2}|\nabla (r B_\phi)|^2\big]}\,,\\
 \beta&=&\frac{-(\mu^2+d_e^2)\Gc'^2}{\rho^2 r^2
\big[c_s^2-v_p^2-\frac{d_e^2}{\rho^2r^2}|\nabla (r B_\phi)|^2\big]}\,,
\end{eqnarray}
 where $c_s^2:=\rho h'(\rho)=c_{is}^2+c_{es}^2$ the Alfv\'en normalized speed of sound, leading eventually to 
\begin{eqnarray}
&&\frac{(\gamma^2+d_e^2)(\mu^2+d_e^2)\Fc'^2\Gc'^2\left[|\nabla\varphi|^2|\nabla\xi|^2-(\nabla\varphi\cdot\nabla\xi)^2\right]}{\rho^4r^4\left(v_p^2+\frac{d_e^2}{\rho^2 r^2}|\nabla(rB_\phi)|^2-c_s^2\right)^2}\nn\\
&&\hspace{1cm}+\frac{1}{1-\left(v_p^2+\frac{d_e^2}{\rho^2 r^2}|\nabla(rB_\phi)|^2\right)/c_s^2}>0\,. \label{ellipticity_final}
\end{eqnarray}
This is the ellipticity condition for the complete system of axisymmetric XMHD equilibrium equations. We observe that since the first term is always non-negative a sufficient (but not necessary) condition for ellipticity is 
\begin{equation}
v_p^2+\frac{d_e^2}{\rho^2 r^2}|\nabla(rB_\phi)|^2<c_s^2\,.\label{ellipticity_simple}
\end{equation}
Observe in \eqref{ellipticity_final} that setting $d_e=0$, i.e.\  neglecting electron inertia, we recover  the Hall MHD ellipticity condition $v_p^2<c_s^2$.

In \cite{Kaltsas2017} it became clear that the XMHD formalism suggests a different kind of static equilibria. Consider a situation where the electron and ion fluids are not static, nevertheless there is no macroscopic mass flow, because they are moving in such a way that the total flow $\bv=(m_i\bv_i+m_e\bv_e)/(m_i+m_e)$ vanishes everywhere. So,  if we assume $\bv\equiv 0$  in  expression \eqref{v_phi},  we conclude that $\varphi=\xi$. Thus \eqref{ellipticity_final} reduces to
\begin{equation}
\frac{d_e^2}{\rho^2 r^2}|\nabla(rB_\phi)|^2<c_s^2\,. \label{ellipticity_static}
\end{equation}
Therefore, in principle, elliptic-hyperbolic transitions are possible even for zero macroscopic flow, something that cannot happen within the framework of the MHD and HMHD.  This is indeed plausible because  static XMHD equilibrium does not mean strictly static ions and electrons -- if that were the case, there would be no current at all. However, we need to clarify that the violation of the ellipticity condition \eqref{ellipticity_static} would require rather peculiar conditions, i.e. very high current density, since $|\nabla(rB_\phi)|^2/r^2$ is the poloidal current density squared and very low mass density, because the speed of sound decreases, if for example a polytropic equation of state is adopted $(p\propto\rho^\Gamma)$, while the lhs of \eqref{ellipticity_static} increases with density decrease. Therefore, a transition to the hyperbolic regime requires sufficiently small mass density and sufficiently high poloidal current density, which effectively means that the difference of the poloidal electron and ion velocities is large enough. This can be seen by rewriting Eq.~\eqref{ellipticity_static} as $(m_e/m)|\bv_{ip}-\bv_{ep}|^2<\Gamma (mn)^{(\Gamma-1)}$. It is also noted that  since the current version of XMHD involves an expansion in the smallness of $m_e/m_i$, although the model is self-consistent, a straightforward correspondence between the two-fluid and the XMHD quantities is not absolutely accurate.   

We point out that \eqref{ellipticity_static} holds also for purely toroidal flows ($v_p=0$), because in that case $\varphi=f(\xi)$ (see Eq.~\eqref{v_p}), so again the first term of \eqref{ellipticity_final} vanishes.   Another case that admits a simplified version of the ellipticity condition \eqref{ellipticity_final} is when one of the two free functions $\Fc$, $\Gc$ is constant, say  $\Gc'=0$. In this case poloidal flow is present and the flow surfaces coincide with the level sets of the stream function $\varphi$. 
For $\Gc'=0$ 
Eq.~\eqref{ellipticity_final} reduces to Eq.~\eqref{ellipticity_simple} that represents now both a  necessary and sufficient ellipticity condition.

As a final point we address the  following reasonable question: Why does the more generic case of two-fluid equilibria possesses ellipticity condition simpler in form?  As stated before,  the quasineutrality condition is the source of the  complication, for it causes  the two stream functions to be  related through a single Bernoulli equation. In the two-fluid case there exist two Bernoulli equations for the two mass densities, each one of which contains a dependence on the gradient of the corresponding stream function and each GS equation contains only the corresponding mass density function. As a consequence the principal symbol has only diagonal elements and the ellipticity condition for each fluid becomes trivial because it results from the requirement that all diagonal elements must have no real roots. This requirement leads eventually to the pair of inequalities \eqref{ellipticity_2f} instead of a single inequality. 

Concluding we emphasize that the present work is a companion  piece  to the two previous studies \cite{Kaltsas2017,Kaltsas2018a} on  XMHD equilibria, which were concerned with the derivation of the equilibrium equations using Hamiltonian structure. Those equations are new in the literature and, therefore, their properties are not yet elucidated. One feature of particular importance is the classification of the equilibrium PDEs. We examined this problem by deriving ellipticity conditions for the complete system and by further examining some special cases. It turned out that the quasineutrality assumption together with the inclusion of electron inertia are of importance for the final form of the ellipticity condition. We deduced a sufficient condition, which becomes necessary under certain assumptions, indicating that  electron inertia lowers  the threshold of the maximum poloidal center of mass velocity for the system to remain elliptic.  In particular, the electron inertial contribution may become considerable within regions of low mass density. Also,  we arrived at the interesting conclusion that in the context of XMHD even static equilibrium equations can become hyperbolic, a consequence of the quasineutrality condition and electron inertia. 

This work has been carried out within the framework of the EUROfusion Consortium and has received funding from the National Programme for the Controlled Thermonuclear Fusion, Hellenic Republic. The views and opinions expressed herein do not necessarily reflect those of the European Commission. D.A.K. was financially supported by the General Secretariat for Research and Technology (GSRT) and the Hellenic Foundation for Research and Innovation (HFRI). P.J.M. was supported by the US Department of Energy contract DE-FG05-80ET-53088. A portion of this material is based upon work supported by the National Science Foundation under Grant No. 1440140, while PJM was in residence at the Mathematical Sciences Research Institute in Berkeley, California, during the Fall of 2018.

\begin{appendices}
\renewcommand\thetable{\thesection\arabic{table}}
\renewcommand\thefigure{\thesection\arabic{figure}}

\medskip

\end{appendices}

\end{document}